\documentclass[prb,twocolumn,showpacs,preprintnumbers,float]{revtex4-1}

\usepackage[colorlinks=true,urlcolor=blue,citecolor=blue,linkcolor=blue,breaklinks=true]{hyperref}
\usepackage[dvipdfmx]{graphicx}
\usepackage{amsmath,amssymb}
\usepackage{times}
\usepackage{physics}
\usepackage{float}
\usepackage{xcolor}
\usepackage{colortbl}
\usepackage{tabularx}
\usepackage{ulem}
\usepackage{multirow}
\usepackage{bm}
\usepackage{xr}
\usepackage{comment}
\usepackage{cleveref}
\usepackage{siunitx}
\usepackage{ulem}
\usepackage[utf8]{inputenc}

\definecolor{mygreen}{rgb}{0.0, 0.8, 0.2}

\definecolor{caribbeangreen}{rgb}{0.0, 0.8, 0.6}

\newcommand{\ul}[1]{#1 *}

\newcommand{\cg}[1]{\multicolumn{1}{>{\columncolor[gray]{0.85}}>{\centering}p{10.5mm}|}{#1}} 
\newcommand{\cgf}[1]{\multicolumn{1}{|>{\columncolor[gray]{0.92}}>{\centering}p{11mm}|}{#1}}
\newcolumntype{F}{>{\columncolor[gray]{0.85}}>{\centering}p{11mm}} 
\newcolumntype{L}{>{\centering}p{10.5mm}} 

\begin{document}

\title{Magnetic interactions in intercalated transition metal dichalcogenides: a study based on {\it ab initio} model construction}
\author{Tatsuto Hatanaka$^1$}
\email{hatanaka-tatsuto346@g.ecc.u-tokyo.ac.jp}
\author{Takuya Nomoto$^2$}
\email{nomoto@ap.t.u-tokyo.ac.jp}
\author{Ryotaro Arita$^{2,3}$}
\email{arita@riken.jp}
\affiliation{$^1$ Department of Applied Physics, The University of Tokyo, Hongo, Bunkyo-ku, Tokyo, 113-8656, Japan \\
$^2$ Research Center for Advanced Science and Technology, University of Tokyo, Komaba Meguro-ku, Tokyo 153-8904, Japan \\
$^3$ RIKEN Center for Emergent Matter Science (CEMS), Wako 351-0198, Japan}
\date{\today}

\begin{abstract}
Transition metal dichalcogenides (TMDs) are known to have a wide variety of magnetic structures by hosting other transition metal atoms in the van der Waals gaps. To understand the chemical trend of the magnetic properties of the intercalated TMDs, we perform a systematic first-principles study for 48 compounds with different hosts, guests, and composition ratios. Starting with calculations based on spin density functional theory, we derive classical spin models by applying the Liechtenstein method to the {\it ab initio} Wannier-based tight-binding model. We show that the calculated exchange couplings are overall consistent with the experiments. In particular, when the composition rate is 1/3, the chemical trend can be understood in terms of the occupation of the 3$d$-orbital in the intercalated transition metal. The present results give us a useful guiding principle to predict the magnetic structure of compounds that are yet to be synthesized.
\end{abstract}

\maketitle

\section{Introduction}\label{intro}
Transition metal dichalcogenides (TMDs) are two-dimensional layered materials of
the type $TX_2$, where $T$ is a transition metal atom, and $X$ is a chalcogen atom. They
offer a fascinating playground to study various physical phenomena such as unconventional superconductivity, exotic charge density waves, emerging spin, valley, and exciton physics~\cite{TMDreview, TMDreview2, TMDreview3, TMDreview4}. 
One of their characteristic features in bulk and thin films with atomic-scale thickness is that they can serve as an intercalation host. Namely, various guest elements can be accommodated in the van der Waals (vdW) gaps between each layer of $TX_2$, changing the physical properties of the system dramatically. In particular, when 3$d$ transition metal atoms ($M$) are intercalated, a variety of magnetic states such as helical spin states~\cite{MORIYA1982209,doi:10.1063/1.4939558,doi:10.1063/1.4896950,PhysRevLett.108.107202,KOUSAKA2009250,CrTa3S6_chiral}, half-metallic states~\cite{PhysRevB.68.134407}, noncollinear antiferromagnetic states~\cite{Parkin_1983,PhysRevB.92.214419}, anisotropic in-plane ferromagnetic states~\cite{Yamasaki_2017,TaTa6Se12} emerges, for which intriguing transport phenomena such as the anomalous Hall effect ~\cite{CoTa3S6-AHE,FeTa3S6,PhysRevB.77.014433} and crystalline Hall effect ~\cite{doi:10.1126/sciadv.aaz8809} have been investigated intensively.

It is an interesting question whether such various magnetic states and properties realized in the intercalated TMDs can be reproduced from first principles and described/understood in terms of a simple model. It is also a non-trivial challenge to predict unknown magnetic properties for compounds that are yet to be synthesized. For these problems, recently, several {\it ab initio} studies have been performed. For example, a calculation based on density functional theory (DFT) has successfully shown that the most stable state in $\mathrm{\textit{M}_{1/3}NbS_{2}}$ where $M$=(Fe, Co) has a noncoplanar magnetic structure for which the topological Hall effect is expected to be observed~\cite{PhysRevMaterials.6.024201}. 
For $M$=(Cr, Mn, Fe), effective spin models were derived from first principles, and the origin of the characteristic helical magnetic structure has been discussed~\cite{PhysRevB.94.184430}. 
However, the general chemical trend of the host- and guest-dependence of the magnetic property of the intercalated TMDs is yet to be fully understood, and a systematic study for various host $TX_2$ and guest $M$ with different composition ratios is highly desired.

To determine the most stable magnetic structure for a given material, there are several established approaches. One is of course a calculation based on spin DFT (SDFT), which usually works successfully for transition metal compounds~\cite{Mita}. However, this approach is numerically expensive and not so efficient when the magnetic unit cell is large.
Another promising approach is deriving a classical spin model from SDFT calculation for a magnetic state (typically the ferromagnetic state) for which the numerical cost is not so expensive.
Once a classical spin model is derived, we can determine the stable magnetic structures even when the magnetic unit cell is large.

The local force method, equivalently called the Liechtenstein formula~\cite{Liechtenstein_1984}, is often used to construct such effective spin models. With this method,
we can evaluate the exchange interactions in the spin model by estimating the energy change against spin rotations.
This formula has been successfully applied to the calculations for the magnetic transition temperatures of transition metals ~\cite{doi:10.1143/JPSJ.68.620}, noncollinear magnets, and magnetic alloys ~\cite{doi:10.1143/JPSJ.69.3072}.  
While it was originally formulated for the multiple scattering theory with the Green's functions and implemented in SDFT calculations with the Korringa-Kohn-Rostoker (KKR) theory,
it is applicable to the tight-binding model based on {\it ab initio} Wannier functions~\cite{PhysRevB.102.014444,PhysRevLett.125.117204,PhysRevResearch.2.043144}.

In this study, we first performed a systematic SDFT calculation for $M_{x}TX_{2}$ where $M$ = (V, Cr, Mn, Fe, Co, Ni), $T$ = (Nb, Ta), and $X$ = (S, Se) with $x$ = $1/3$ and $1/4$ (48 compounds in total).
Starting with the calculations for the representative ferromagnetic state of Cr$_{x}TX_{2}$, we construct classical spin models by applying the local force method to the Wannier-based tight-binding model.
We then determine the most stable magnetic structure for each material by examining the sign of the exchange interactions. In this approach, we discuss the possibility of the intra-layer AF states which are numerically expensive to investigate by SDFT calculation.
We show that the theoretical results agree well with the magnetic orders experimentally reported. Moreover, we find for $x=1/3$ compounds that a simple model can give a unified explanation for the material dependence of the stable spin configuration in terms of the filling of the 3$d$ orbitals of the intercalated transition metals. This observation gives us a useful guiding principle to predict magnetic properties of intercalated TMDs which are yet to be synthesized.

\section{Method}
Starting from the DFT calculation, we first construct a tight-binding Hamiltonian based on the Wannier function. We then applied the Liechtenstein formula~\cite{Liechtenstein_1984} and derived an effective spin model.
We neglect the spin-orbit coupling effect so that spin canting due to the Dzyaloshinsky-Moriya interaction~\cite{DZYALOSHINSKY1958241,DMMoriya} is not considered in the present study.
Then, the sum of effective interactions ($J_{0}$) and each interaction ($J_{ij}$) are perturbatively evaluated by rotating a spin from the ferromagnetic state and examining the changes of the total energy.

Let us first consider the classical Heisenberg Hamiltonian: 
\begin{eqnarray}
H_{\textrm{s}} = -2\sum_{\langle i,j\rangle}J_{ij}\bm{s}_{i}\cdot\bm{s}_{j}
\end{eqnarray}
We then introduce
$\delta E_{i}$ as the energy change when we rotate the spin at site $i$ by $\theta_{i}$ from the ferromagnetic state, and $\delta E_{ij}$ as the energy change when 
the spin at site $j$ is also rotated by $\theta_{j}$ on the same rotation axis. It should be noted that
$\delta E_{i}$ and $\delta E_{ij}$ are directly related with 
$2\sum_{j\neq i}J_{ij}$ and $-2J_{ij}$, respectively:
\begin{eqnarray}\label{eq:model}
\begin{split}
\pdv[2]{\delta E_{i}}{\theta_{i}} &= 2\sum_{j\neq i}J_{ij} \\
\pdv[2]{\delta E_{ij}}{\theta_{i}}{\theta_{j}} &= -2J_{ij}
\end{split}
\end{eqnarray}

Next, we consider the tight-binding Hamiltonian defined as follows,
\begin{eqnarray}
H_{\textrm{TB}} = \sum_{\langle i,j\rangle}A_{ij}c^{\dagger}_{i}c_{j}
\end{eqnarray}
where the indices $i,j$ run over all degrees of freedom that specify the Wannier functions, namely, lattice vectors, sublattices, atomic or molecular orbitals, and spins.
Using the Green's function for the tight-binding Hamiltonian, we can calculate the energy change due to the spin rotation.
In the Green's functions formalism, the free energy $F$ of the system (\ref{F_Green}) is expressed as,
\begin{eqnarray}\label{F_Green}
F = -T \sum_{\omega_{n}}e^{i\omega_{n}0^{+}}\textrm{Tr ln}[G^{-1}(i\omega_{n})].
\end{eqnarray}
where $\omega_{n} = (2n+1)\pi/\beta$ denotes the electronic Matsubara frequency and the Green's function $G$ is given by $G(i\omega_{n})^{-1}=(i\omega_{n}\delta_{ij}-A_{ij})$.
If we rotate the spins as in the case of the Heisenberg model, the changes in the free energy, i.e., $\delta F_{i}$ and $\delta F_{ij}$, are given by the following equations:
\begin{eqnarray}
\pdv[2]{\delta F_{i}}{\theta_{i}} &=& -2T\sum_{\omega_{n}}\textrm{Tr}_{jl\sigma}\qty[B_{i}G^{\uparrow}_{ij}B_{j}G^{\downarrow}_{ji}] \nonumber \\
&&+2T\sum_{\omega_{n}}\textrm{Tr}_{l\sigma}\qty[B_{i}G^{\uparrow}_{ii}B_{i}G^{\downarrow}_{ii}] \\
\pdv[2]{\delta F_{ij}}{\theta_{i}}{\theta_{j}} &=& T\sum_{\omega_{n}}\textrm{Tr}_{l\sigma}\left[B_{i}G^{\uparrow}_{ij}B_{j}G^{\downarrow}_{ji}\right]
\end{eqnarray}
where $B$ stands for the effective magnetic field, 
namely, the spin splitting in the calculation based on the local spin density approximation (LSDA). 
By comparing these expressions with equation (\ref{eq:model}), we can evaluate $J_{ij}$ for the itinerant tight-binding Hamiltonian (eq.~(\ref{F_Green})). 

\section{Computational Details}
\begin{figure}[htbp]
\centering
\includegraphics[keepaspectratio, scale=0.55]{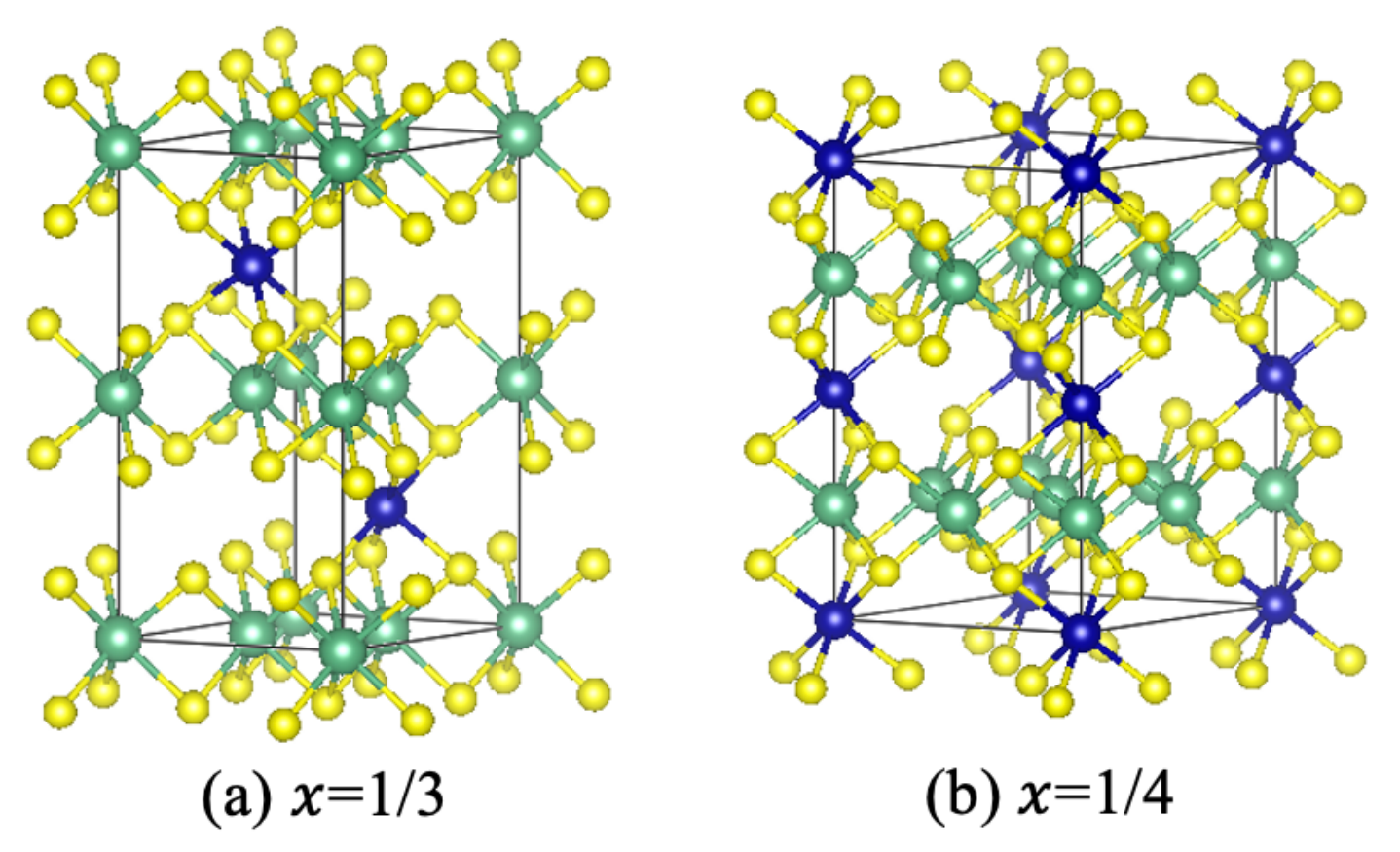}
\caption{Crystal Structures of intercalated TMDs $M_xTX_2$ (a) for $x=1/3$ and (b) for $x=1/4$. 
Green, yellow and blue spheres represent $T$, $X$ and $M$ elements, respectively. \label{structure}}
\end{figure}

\subsection{DFT calculation}
We used the Vienna \textit{Ab initio} Simulation Package code~\cite{PhysRevB.54.11169} for SDFT calculations of intercalated TMDs.
The Perdew–Burke-Ernzerhof  exchange-correlation functional ~\cite{PhysRevLett.77.3865} and the projector augmented wave method ~\cite{PhysRevB.50.17953,PhysRevB.59.1758} were used.

We show the crystal structures of intercalated TMDs in Fig.~\ref{structure}. There are two intercalated transition metals per unit cell, which are located in different vdW gaps. We see that the intercalated transition metals are surrounded by a distorted octahedron formed by chalcogen atoms. We can also see that intercalated transition metals form a hexagonal close-packed lattice when $x=1/3$ and a triangular lattice stacked along the $c$-axis when $x=1/4$.

We performed structural optimization for all target compounds. In this optimization, we assumed that the spin configuration is ferromagnetic (FM), and the lattice parameters and internal coordinates were optimized, keeping the original space group symmetries $P6_322$ for $x=1/3$ and $P6_3/mmc$ for $x=1/4$. 
For materials having no magnetization in the SDFT calculations, we performed SDFT+$U$ calculations.
The value of $U$ was set as $U=3$ eV for Fe$_{1/4}$XSe$_2$ ($X=$ Nb and Ta), the Co-, and Ni-intercalated compounds.
As mentioned before, there are two intercalated transition metals per unit cell (see, Fig.~\ref{structure}). In the SDFT calculations, we focus on the magnetic structures that do not expand the unit cell. Thus, we consider only the antiferromagnetic (AFM) state having the interlayer antiferromagnetic and intralayer ferromagnetic structure.
The energies of FM and AFM states were calculated for optimized structures and compared with each other.
The energy cut-off for the plane-wave basis set was set to 500 eV, and a 12$\times$12$\times$8 k-point grid for the primitive cell of the intercalated TMDs was used in the structural optimization and the calculations of the ground state energies.

\subsection{Construction of Wannier-based tight-binding model}
\begin{figure}[htbp]
  \includegraphics[keepaspectratio, scale=0.4]{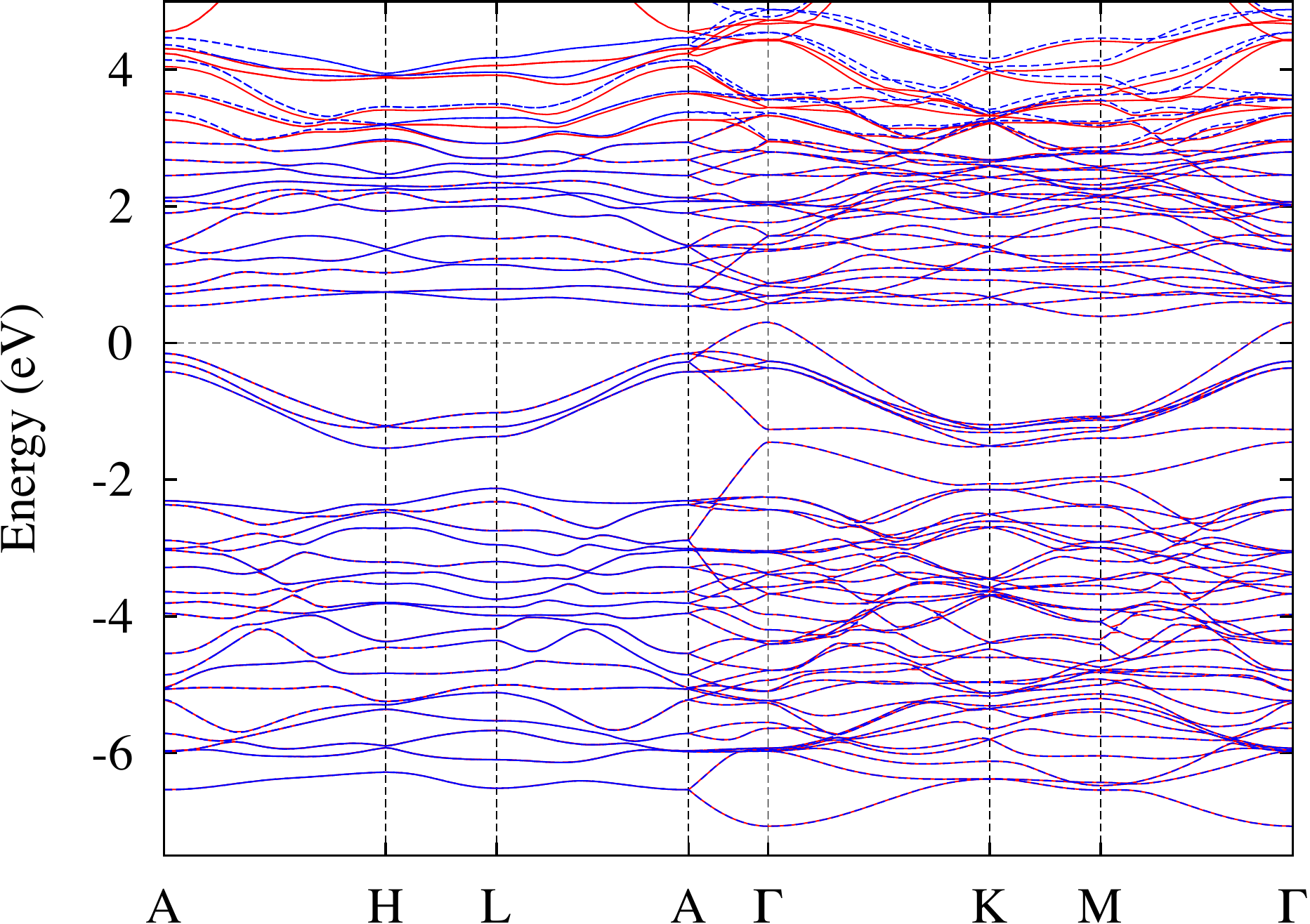}
  \includegraphics[keepaspectratio, scale=0.4]{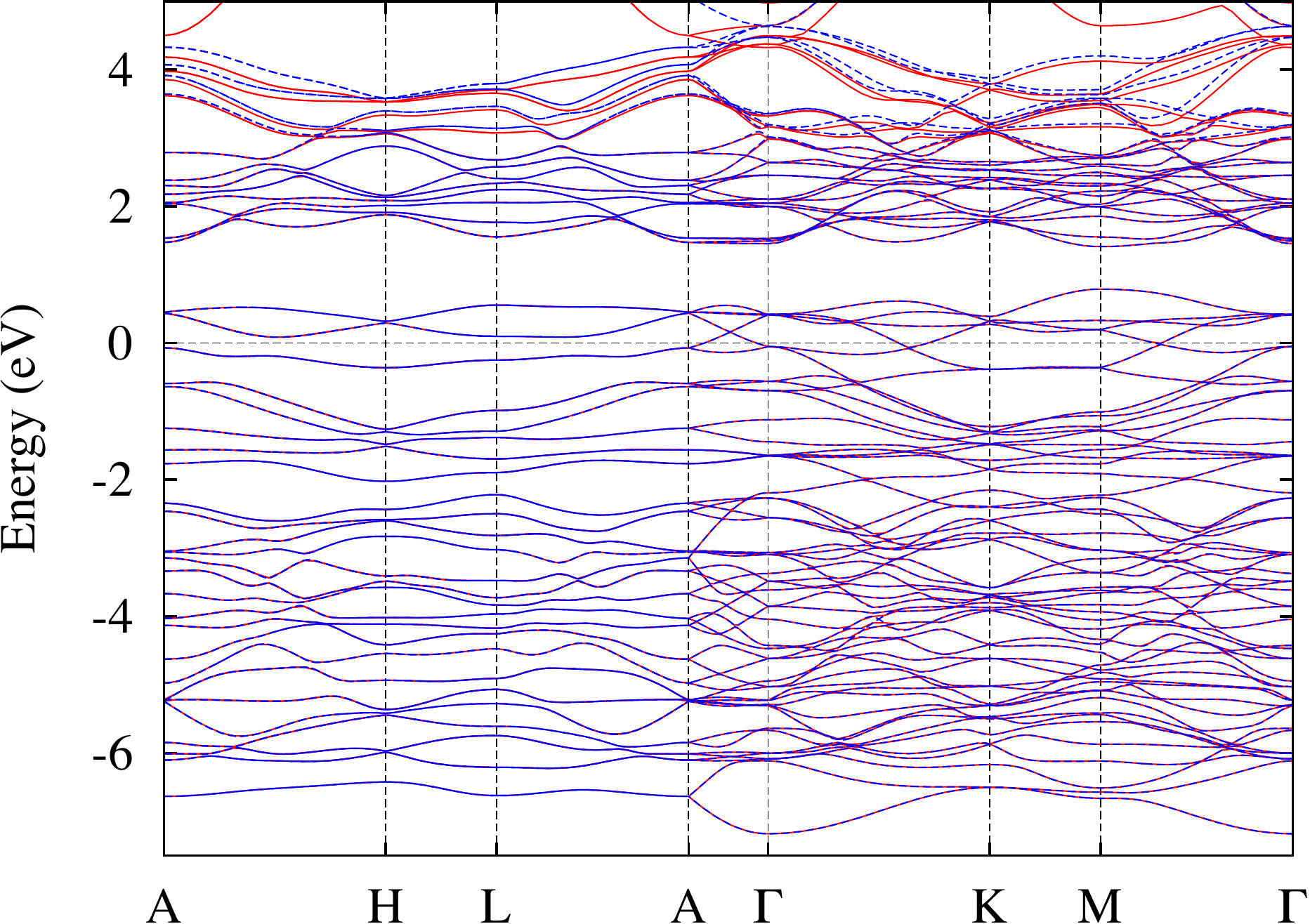}
  \caption{Band structures of (upper) majority spin and (lower) minority spin of $\mathrm{Cr_{1/3}NbS_{2}}$ in the ferromagnetic state. The energy are measured from the Fermi level. Blue lines are calculated from DFT calculations, and red lines are from the Wannier functions.\label{band3}}
\end{figure}

Wannier functions were constructed by using the Wannier90 code~\cite{Pizzi2020}.  
In Fig. \ref{band3}, we show the band structures of $\mathrm{Cr_{1/3}NbS_{2}}$ as a representative example.
The inner window to fix the low energy band dispersion was set from -8 to  2 eV.
The energy cut-off for the plane-wave basis was set to 500 eV. A 12$\times$12$\times$8 $k$-point grid was used in calculating FM reference states,
and a 6$\times$6$\times$4 sampling $k$-point grid was used for constructing Wannier functions.

\subsection{Evaluation of exchange interactions}\label{IIIC}
We applied the Liechtenstein formula to the Cr-intercalated compounds, and the exchange interactions for the other transition metal-intercalated compounds were evaluated by shifting the Fermi level, which corresponds to the rigid band approximation. 
According to the results of the DFT calculation, only intercalated transition metals have a sizeable magnetic moment, and those of the other atoms are negligibly small in the FM order. 
Thus we ignored interactions other than those between intercalated transition metals, and extracted the spin model, whose interactions are finite only between intercalated transition metals. 

A 8$\times$8$\times$8 k-point grid was used 
in the evaluation of Eqs. (5) and (6). 
Inverse temperature $\beta$ was set to 500 $\textrm{eV}^{-1}$.
In order to reduce the computational cost, we use the intermediate representation of the Green's function~\cite{PhysRevB.96.035147,CHIKANO2019181} in the Liechtenstein formula.

\section{Results and Discussion}
\subsection{Stable magnetic order according to DFT calculation \label{sec:dft}}
We first summarize the experimentally observed magnetic structures in  Table~\ref{main}(a). Table~\ref{main}(b) shows the results of the DFT calculations, where we compare the energies of the FM and AFM states.
We can see from Table~\ref{main}(a) and \ref{main}(b) that the experimental magnetic structures in 16 out of the 23 compounds are successfully reproduced in the DFT calculations. The remaining 6 compounds except for $\mathrm{Fe_{1/4}TaS_{2}}$ (namely, $\mathrm{Co_{1/3}NbS_{2}}$, $\mathrm{Co_{1/3}TaS_{2}}$, $\mathrm{Cr_{1/4}NbS_{2}}$, $\mathrm{Cr_{1/4}NbSe_{2}}$, $\mathrm{Mn_{1/4}NbSe_{2}}$, and $\mathrm{Fe_{1/4}NbSe_{2}}$) are known to be AFM in the experiments but predicted to be FM in the DFT calculation. It should be noted that we did not consider intralayer AFM states in the DFT calculation because the magnetic unit cell becomes too large. Thus the intralayer magnetic structure is always FM, and only the interlayer magnetic structure can be AFM. We will see later in Table \ref{main}(c) that we obtain the correct AFM ground states in $\mathrm{Cr_{1/4}NbS_{2}}$  and $\mathrm{Cr_{1/4}NbSe_{2}}$ based on the spin model calculations derived by the Liechtenstein formula. On the other hand, in the case of $\mathrm{Fe_{1/4}TaS_{2}}$, the AFM state is more stable than the FM state in the DFT calculation, while it is FM in the experiment. We will discuss this discrepancy in Sec.~\ref{exchange}. 

\begin{table*}[htbp]
  \begin{tabular}{lcc}
    &$x$=1/3 & $x$=1/4 \\
    \\
    (a)\ \ Exp. &
    \begin{minipage}{0.45\linewidth}
      \begingroup
      \renewcommand{\arraystretch}{1.6}
      \begin{tabular}{|F|L|L|L|L|L|L|r}
        \cline{1-7}
        \cgf{Exp.}&\cg{V}&\cg{Cr}&\cg{Mn}&\cg{Fe}&\cg{Co}&\cg{Ni}&\\
        \cline{1-7}
        $\mathrm{NbS_{2}}$&AF\cite{PhysRevB.103.174431,PhysRevMaterials.4.054416} &F\cite{HM-FM}&F\cite{HM-FM}&AF\cite{PhysRevX.12.021003,FeNb3S6}&AF\cite{PhysRevB.105.155114,Parkin_1983,CoNb3S6}&AF\cite{Mn-NiNb3S6}&\\
        \cline{1-7}
        $\mathrm{NbSe_{2}}$&-&F\cite{doi:10.1063/1.4939558}&-&AF\cite{eve2012structural}&-&-&\\
        \cline{1-7}
        $\mathrm{TaS_{2}}$&AF\cite{PhysRevMaterials.4.054416}&F\cite{Yamasaki_2017,HM-FM}&F\cite{MnTa3S6}&F\cite{FeTa3S6}&AF\cite{Parkin_1983}&AF\cite{doi:10.1080/13642818008245371}&\\
        \cline{1-7}
        $\mathrm{TaSe_{2}}$&-&-&-&-&-&-&\\
        \cline{1-7}
      \end{tabular}
      \endgroup
    \end{minipage}&
    \begin{minipage}{0.45\linewidth}
      \begingroup
      \renewcommand{\arraystretch}{1.6}
      \begin{tabular}{|F|L|L|L|L|L|L|r}
        \cline{1-7}
        \cgf{Exp.}&\cg{V}&\cg{Cr}&\cg{Mn}&\cg{Fe}&\cg{Co}&\cg{Ni}&\\
        \cline{1-7}
        $\mathrm{NbS_{2}}$&-&AF~\cite{VANLAAR1971154}&F\cite{doi:10.1143/JPSJ.55.347,MnNb4S8}&AF\cite{doi:10.1080/01418638108222164,TSUJI2001213}&-&-& \\
        \cline{1-7}
        $\mathrm{NbSe_{2}}$&-&AF\cite{VOORHOEVEVANDENBERG1971167}&AF\cite{doi:10.1143/JPSJ.55.347}&AF\cite{VOORHOEVEVANDENBERG1971167,doi:10.1080/13642818008245370}&-&-& \\
        \cline{1-7}
        $\mathrm{TaS_{2}}$&-&-&F\cite{doi:10.1143/JPSJ.55.347,doi:10.1080/13642818008245370}&F\cite{PhysRevB.77.014433,PhysRevB.75.104401,PhysRevLett.107.247201}&-&-& \\
        \cline{1-7}
        $\mathrm{TaSe_{2}}$&-&-&-&-&-&F\cite{NiTa4Se8}& \\
        \cline{1-7}
      \end{tabular}
      \endgroup
    \end{minipage}\label{tab:exp} \\ \\
    (b)\ \ DFT &
    \begin{minipage}{0.45\linewidth}
      \begingroup
      \renewcommand{\arraystretch}{1.6}
      \begin{tabular}{|F|L|L|L|L|L|L|r}
        \cline{1-7}
        \cgf{DFT}&\cg{V}&\cg{Cr}&\cg{Mn}&\cg{Fe}&\cg{Co}&\cg{Ni}&\\
        \cline{1-7}
        $\mathrm{NbS_{2}}$&AF&F&F&AF&\ul{F}&AF&\\
        \cline{1-7}
        $\mathrm{NbSe_{2}}$&AF&F&F&AF&AF&AF&\\
        \cline{1-7}
        $\mathrm{TaS_{2}}$&AF&F&F&F&\ul{F}&AF&\\
        \cline{1-7}
        $\mathrm{TaSe_{2}}$&F&F&F&AF&AF&AF&\\
        \cline{1-7}
      \end{tabular}
      \endgroup
    \end{minipage} &
    \begin{minipage}{0.45\linewidth}
      \begingroup
      \renewcommand{\arraystretch}{1.6}
      \begin{tabular}{|F|L|L|L|L|L|L|r}
        \cline{1-7}
        \cgf{DFT}&\cg{V}&\cg{Cr}&\cg{Mn}&\cg{Fe}&\cg{Co}&\cg{Ni}&\\
        \cline{1-7}\cline{1-7}
        $\mathrm{NbS_{2}}$&F&\ul{F}&F&AF&AF&F& \\
        \cline{1-7}
        $\mathrm{NbSe_{2}}$&F&\ul{F}&\ul{F}&\ul{F}&AF&F& \\
        \cline{1-7}
        $\mathrm{TaS_{2}}$&F&F&F&\ul{AF}&AF&F& \\
        \cline{1-7}
        $\mathrm{TaSe_{2}}$&F&F&F&F&AF&F& \\
         \cline{1-7}
      \end{tabular}
      \endgroup
    \end{minipage}\label{tab:dft}\\ \\
    (c)\ \ Model &
    \begin{minipage}{0.45\linewidth}
      \begingroup
      \renewcommand{\arraystretch}{1.6}
      \begin{tabular}{|F|L|L|L|L|L|L|r}
        \cline{1-7}
        \cgf{Model}&\cg{V}&\cg{Cr}&\cg{Mn}&\cg{Fe}&\cg{Co}&\cg{Ni}&\\
        \cline{1-7}
        $\mathrm{NbS_{2}}$&AF&F&F/AF&AF&AF&\ul{F}&\\
        \cline{1-7}
        $\mathrm{NbSe_{2}}$&AF&F&F&AF&AF&F&\\
        \cline{1-7}
        $\mathrm{TaS_{2}}$&AF&F&F&\ul{AF}&AF&\ul{F}&\\
        \cline{1-7}
        $\mathrm{TaSe_{2}}$&F/AF&F&F/AF&AF&AF&F&\\
        \cline{1-7}
     \end{tabular}
     \endgroup
    \end{minipage} &
    \begin{minipage}{0.45\linewidth}
      \begingroup
      \renewcommand{\arraystretch}{1.6}
      \begin{tabular}{|F|L|L|L|L|L|L|r}
        \cline{1-7}
        \cgf{Model}&\cg{V}&\cg{Cr}&\cg{Mn}&\cg{Fe}&\cg{Co}&\cg{Ni}&\\
        \cline{1-7}\cline{1-7}
        $\mathrm{NbS_{2}}$&AF&AF&F/AF&AF&AF&AF& \\
        \cline{1-7}
        $\mathrm{NbSe_{2}}$&AF&AF&F/AF&AF&AF&AF& \\
        \cline{1-7}
        $\mathrm{TaS_{2}}$&AF&AF&F/AF&\ul{AF}&AF&AF& \\
        \cline{1-7}
        $\mathrm{TaSe_{2}}$&AF&AF&F/AF&AF&AF&\ul{AF}& \\
         \cline{1-7}
     \end{tabular}
     \endgroup
    \end{minipage}\label{tab:liech}
  \end{tabular}
  \caption{Stable magnetic structure in the (a) experiments, (b) DFT calculations, and (c) classical spin model derived by the Liechtenstein formula. Letters with an asterisk(*) denote that the theoretical results are not consistent with the experimental results in (a). F, HM, and AF stand for ferromagnetic, helimagnetic, and antiferromagnetic structures, respectively.}
  \label{main}
\end{table*}

\subsection{Exchange constant}\label{exchange}

\begin{figure*}[htbp]
\includegraphics[keepaspectratio, scale=0.5]{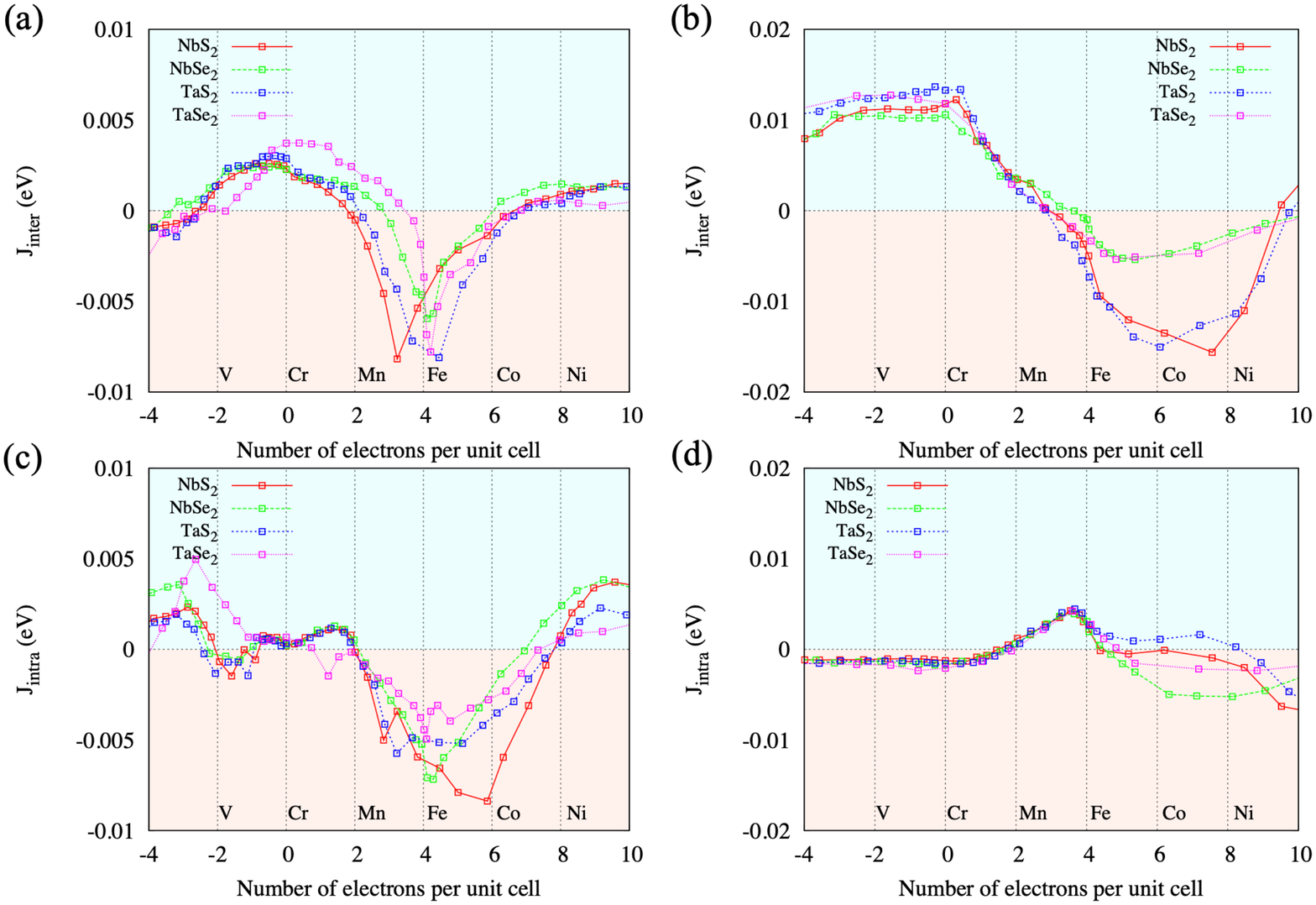}
  \caption{Exchange constants evaluated by the Liechtenstein formula. Figures (a) and (c) are results for $x$=1/3, and (b) and (d) are for $x$=1/4.
  Six vertical black dotted lines correspond to the Fermi level of V-, Cr-, Mn-, Fe-, Co-, and Ni-intercalated compounds, respectively.\label{Jij}}
\end{figure*}

We show the filling (the number of 3$d$ electrons in the unit cell) dependence of the interlayer (Figs.~\ref{Jij}(a), (b)) and intralayer (Figs.~\ref{Jij}(c), (d)) exchange constants  evaluated by the Liechtenstein formula. As we described in Sec.~\ref{IIIC}, we start with the most representative case, i.e., the ferromagnetic state for $M$=Cr. We shift the position of the Fermi level and look at the energy change due to a spin rotation. While we neglect the detail of the host ($TX_2$) dependence on the electronic structure, as we see below, the rigid-band approximation successfully reproduces the overall chemical trend of the experimental results.

Figures \ref{Jij}(a) and \ref{Jij}(c) are the results for $x=1/3$, and 3(b) and 3(d) are those for $x=1/4$. Let us first look at the former. We see that both the intralayer and interlayer interactions have a similar filling dependence. When the number of the 3$d$ electrons is small or large, the exchange constants tend to take a positive small value (FM). On the other hand, when the filling is close to half-filling (as in the cases of Mn, Fe, Co), the interactions tend to be negative (AFM). This result is consistent with the previous study for the Fe- and Co-intercalated $x=1/3$ system in which noncoplanar AFM structures were shown to be favored (Ref.~\onlinecite{PhysRevMaterials.6.024201}) since the hexagonal close-packed lattice is magnetically frustrated when all the nearest neighbor interactions are negative.

Next, let us move on to the case of $x=1/4$. We see that the interlayer exchange constant shown in Fig.~\ref{Jij}(b) does not show a significant host ($TX_2$) dependence for $M$=(V, Cr, Mn, Fe). We see a similar behavior for the intralayer exchange constant (Fig.~\ref{Jij}(d)). Another distinct feature is that the energy scale of the intralayer exchange constant is much smaller than that of the interlayer exchange constant. Namely, the system has a strong coupling along the $c$ axis rather than in the $ab$ plane. 

In Table.~\ref{main}(c), we summarize the stable magnetic structures determined by the sign of the exchange constants. Among 23 compounds for which the magnetic structure is determined experimentally, we can say that the theoretical magnetic structures of 18 compounds are consistent with the experiment. Here, let us note that both the interlayer and intralayer exchange constants change their sign around $M$=Mn. Thus we do not determine which magnetic order is stable for six compounds with $M$=Mn. For $x=1/3$, we have a similar problem for $M$=Ni. Namely, for $\mathrm{Ni_{1/3}NbS_{2}}$ and $\mathrm{Ni_{1/3}TaS_{2}}$, at least one of the intralayer or interlayer interactions is close to zero, indicating that these materials are located near the boundary of the FM and AFM states.

On the other hand, for the case of Fe$_x$TaS$_2$, our approach does not reproduce the experimental results. For $x=1/3$, we should note that while SDFT apparently reproduces the ferromagnetic ground state in the experiment~\cite{FeTa3S6}, the intralayer AFM state is not considered in the calculation. Regarding the reason for the disagreement between theory and experiment, we leave it for future study. 
For $x=1/4$, neither the SDFT calculation nor the Lichtenstein approach reproduces the experimental ferromagnetic ground states. One possible reason is the contribution of the orbital magnetization of the intercalated Fe atoms. While intercalated Fe atoms are shown to have a finite orbital moment of about 33$\%$ of the spin moment [Ref.~\onlinecite{PhysRevLett.107.247201}], the orbital moment is not taken into account in the present calculation.

\subsection{Simple interpretation of the material dependence of the exchange constant for $x=1/3$}\label{mechanism}
\begin{figure}[htbp]
\centering
\includegraphics[keepaspectratio, scale=0.3]{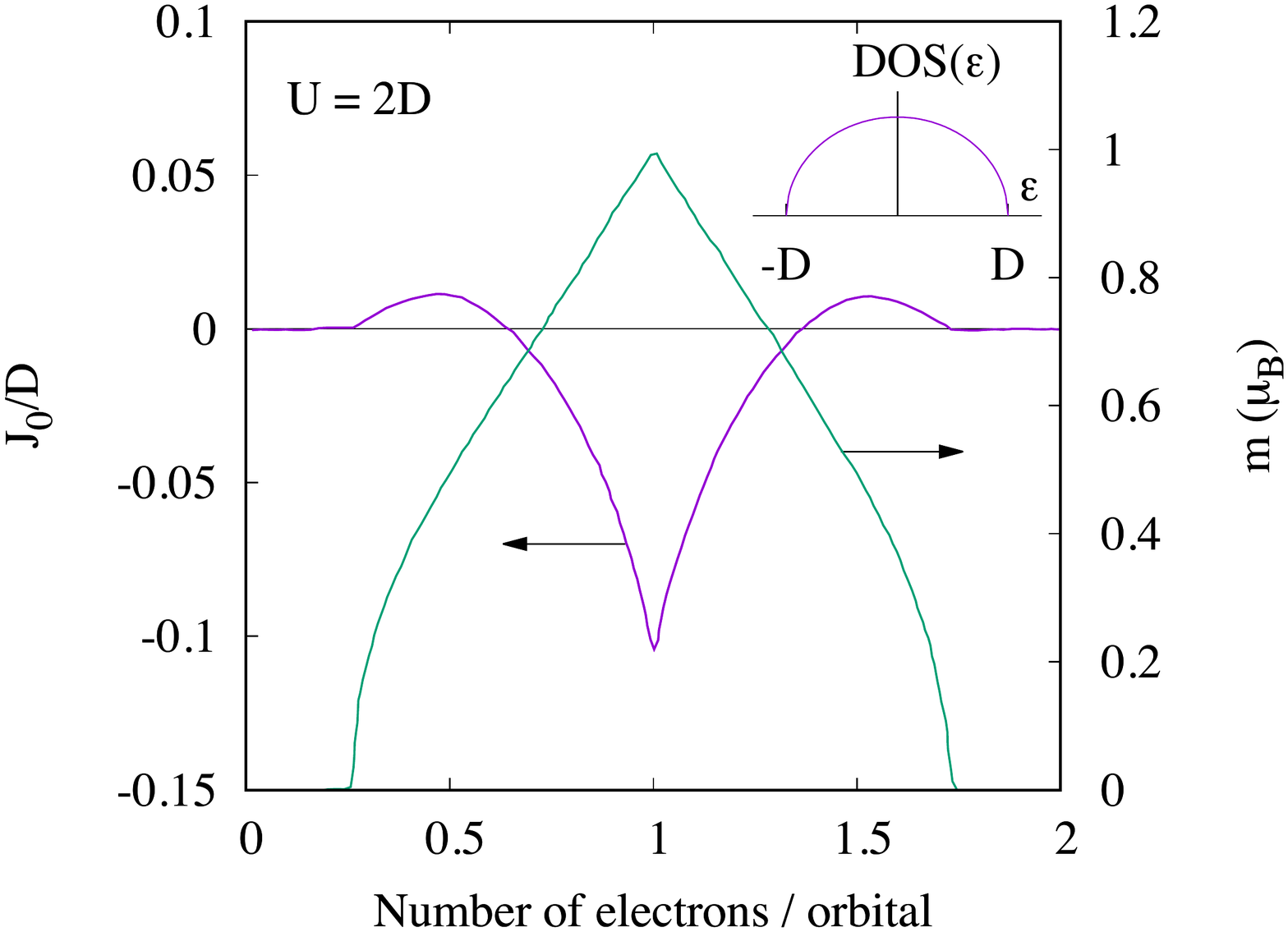}
\caption{Filling dependence of $J_{0}$ for the single orbital Hubbard model on the Bethe lattice. \label{sakuma}}
\end{figure}

As we have seen in Figs.~\ref{Jij}, the energy scale of the interlayer and intralayer exchange constants are similar to each other for $x=1/3$ but very different for $x=1/4$. This result indicates that while the intercalated TMDs are crystallographically two-dimensional, they are magnetically isotropic (three-dimensional) for $x=1/3$ but anisotropic (quasi-one dimensional) for $x=1/4$. In this subsection, let us discuss whether the material dependence of the exchange constant for $x=1/3$ can be understood in terms of a simple single-orbital Hubbard model on the Bethe lattice. When the Coulomb repulsion (the Hubbard $U$) is absent, the system has a semicircular DOS (see the inset of Fig.~\ref{sakuma}). We set the bandwidth $W=2D$ and $U=W$. In Fig.~\ref{sakuma}, we show the filling dependence of $J_{0}(=\sum_{j}J_{ij})$~\cite{800521}. When the filling is closed to 1 (half-filling), the super-exchange mechanism is dominant, and thus $J_0$ takes a negative value. On the other hand, when the filling is very low or high, the double exchange mechanism makes $J_0$ positive. Namely, the system is FM for low and high filling but AFM for half-filling. Interestingly, this behavior can be seen for both the interlayer and intralayer exchange constant for $x=1/4$ (see Figs.~\ref{Jij}(a) and (c)).

\begin{figure}[htbp]
  \centering
  \includegraphics[keepaspectratio, scale=0.5]{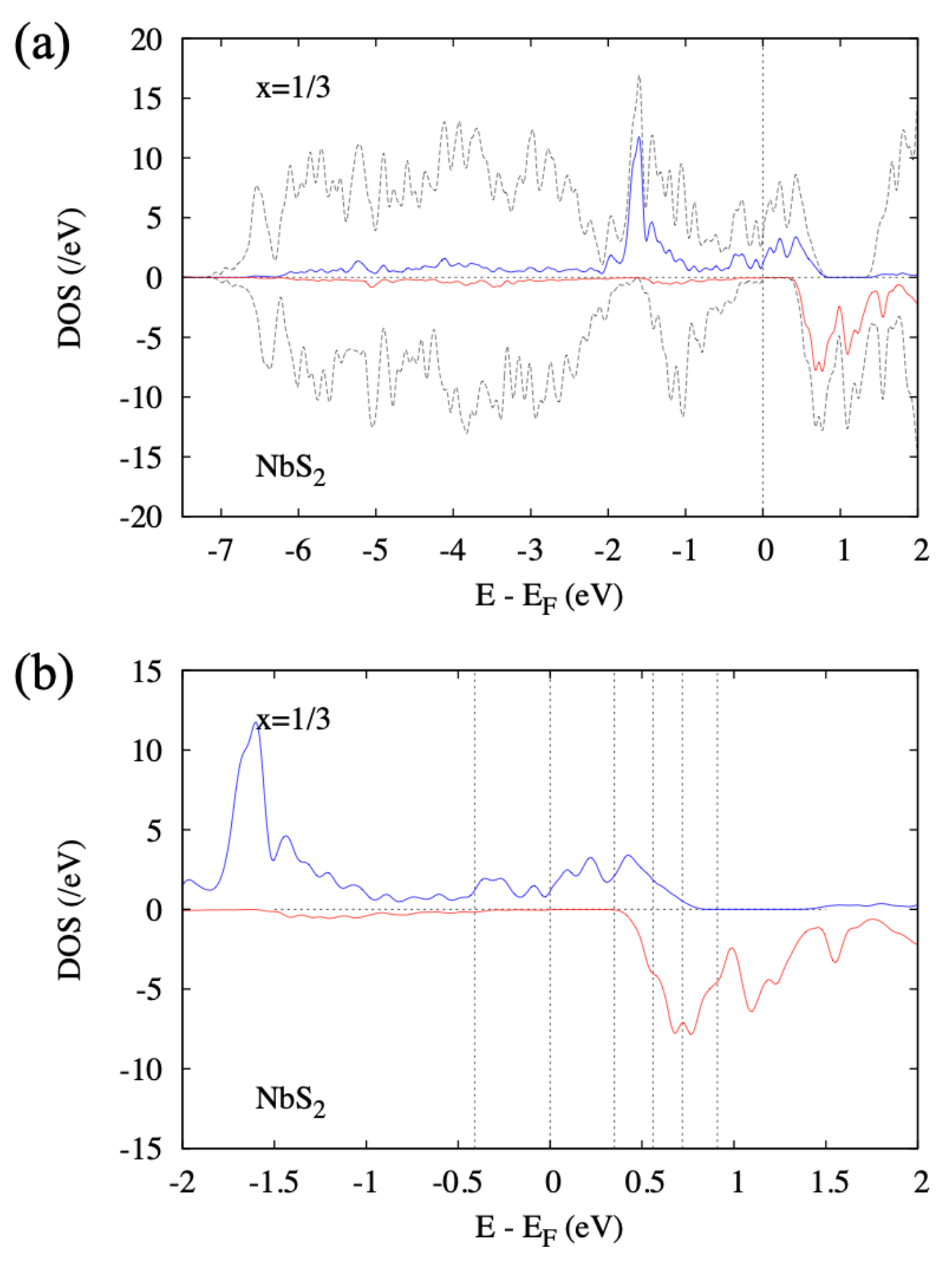}
  \caption{(a) DOS and PDOS of the 3$d$ orbitals for $\textrm{Cr}_{1/3}\textrm{NbS}_{2}$. Black dotted line is DOS and blue (red) line is the PDOS of the 3$d$ orbitals with the majority (minority) spin. (b) Enlarged plot for the PDOS of the 3$d$ orbitals. Six vertical black dotted lines denote the Fermi level of V-, Cr-, Mn-, Fe-, and Ni-intercalated TMDs determined by the rigid band approximation.\label{fig:dos_3}}
\end{figure} 

\begin{figure}[htbp]
\centering
\includegraphics[keepaspectratio, angle=-90, scale=0.3]{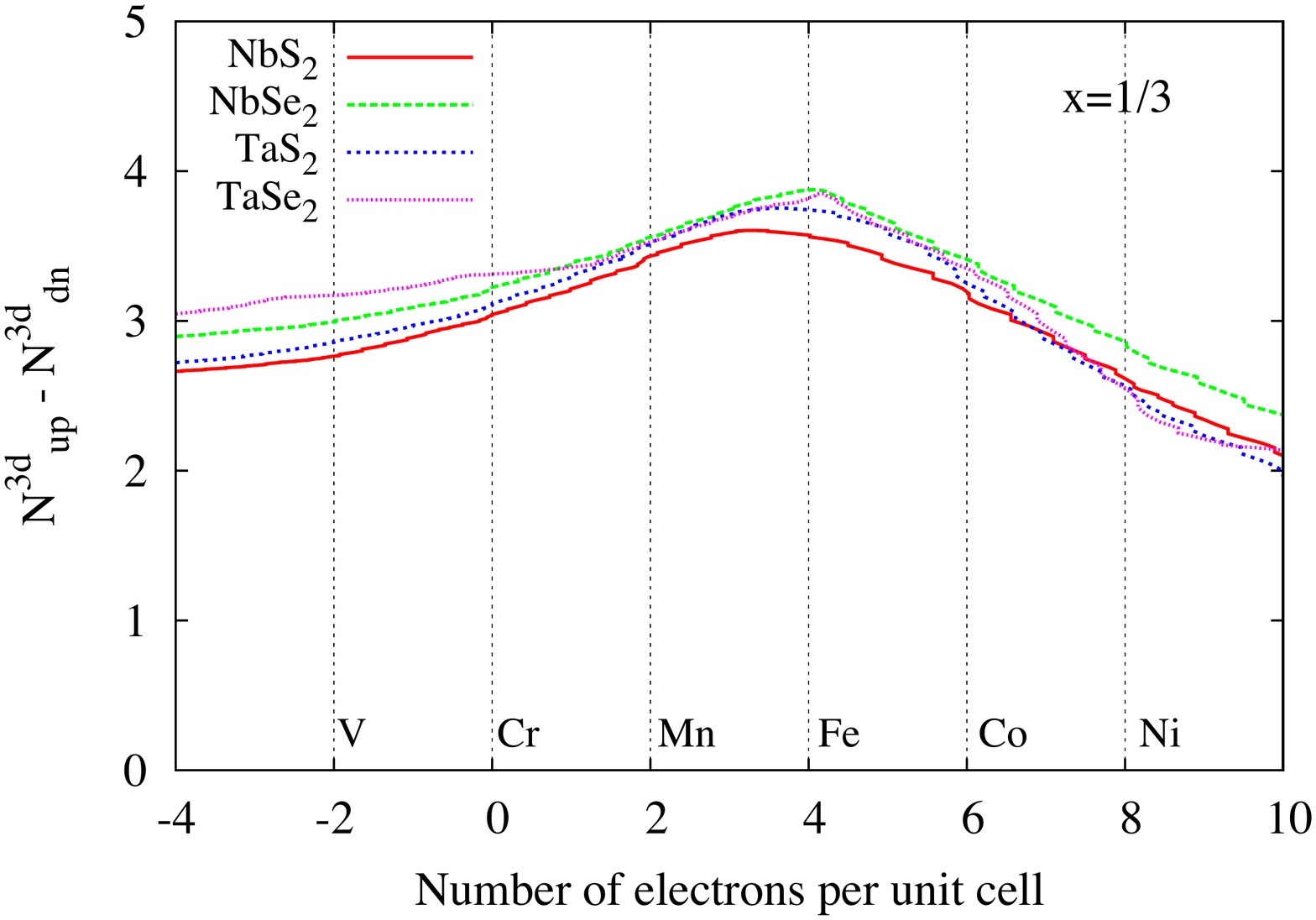}
\caption{Spin polarization (difference between the filling of the majority and minority spins) of the 3$d$ orbitals in $M_{1/3}TX_2$. \label{fill}}
\end{figure}

In Fig.~\ref{sakuma}, we also plot the spin polarization (i.e., the difference between the filling of the majority and minority spins) as a function of the filling. we see that AFM interaction is strongest when the spin polarization is largest. It is interesting to see whether this behavior can also be seen for the exchange constants of $M_{1/3}TX_2$.
As a typical case, let us look into the case of $\textrm{Cr}_{1/3}\textrm{NbS}_{2}$. 
In Fig.~\ref{fig:dos_3}, we show the total DOS and partial DOS (PDOS) of the 3$d$ orbitals.
The vertical black dotted lines in Fig.~\ref{fig:dos_3}(b) denote the Fermi level ($E_F$) for $M$=V, Cr, Mn, Fe, Co, and Ni from the left, respectively. We see that when $E_F=0$ (i.e., the case of $M$=Cr),
the minority spin is almost empty. When $E_F$ is higher than that of Mn ($\sim 0.4$eV), the majority spin starts to be occupied. Thus the spin polarization takes its maximum between $M$=Mn and Fe. To make this situation clearer, in Fig.~\ref{fill}, we plot the spin polarization for $M_{1/3}$NbS$_2$ together with the results for other $M_{1/3}TX_2$.
From these plots, we expect that the AFM interaction becomes strongest for $M$=Mn or Fe, and interestingly again, it is indeed the case seen in Figs.~\ref{Jij}(a) and (c).

\section{Conclusion}
By means of first-principles calculations based on SDFT and {\it ab initio} derivation of the classical spin model based on the Liechtenstein method, we systematically investigated the material dependence of the magnetic interactions in 48 intercalated TMDs $M_xTX_2$, in which a variety of magnetic structures is realized. For both $x$=1/3 and $x$=1/4, our calculations overall succeeded in reproducing the experimental results, and especially for $x$=1/3, we found that the intercalated guest-atom dependence can be simply understood in terms of the filling of the 3$d$ orbitals. The present result will provide a useful guideline to predict magnetic structures in compounds which has not been synthesized. 

\section*{Acknowledgements}
We would like to thank Masaki Nakano for illuminating discussions. We acknowledge the financial support by Grant-in-Aids for Scientific Research (JSPS KAKENHI) Grant No. JP21H04437, JP21H04990 and JP19H05825. T. N. was supported by JST, PRESTO Grant Number JPMJPR20L7, Japan.

\end{document}